\documentclass[submission,copyright,creativecommons]{eptcs}
\providecommand{\event}{TARK 2023} 

\usepackage{iftex}

\ifpdf
  \usepackage{underscore}         
  \usepackage[T1]{fontenc}        
\else
  \usepackage{breakurl}           
\fi

\usepackage{amsfonts}
\usepackage{amssymb}
\usepackage{amsmath}

\newtheorem{theorem}{Theorem}

\newtheorem{definition}{Definition}

\newtheorem{lemma}{Lemma}

\newtheorem{proposition}{Proposition}
\newtheorem{remark}{Remark}

\numberwithin{equation}{section}

\title{Complete Conditional Type Structures
(Extended Abstract)}
\author{Nicodemo De Vito
\institute{Department of Decision Sciences\\
Bocconi University\\
Milan, Italy}
\email{nicodemo.devito@unibocconi.it}
}
\date{May 2023}

\def\titlerunning{Complete Conditional Type Structures}
\def\authorrunning{N. De Vito}

\begin{document}
\maketitle

\begin{abstract}
Hierarchies of conditional beliefs (Battigalli and Siniscalchi 1999) play a
central role for the epistemic analysis of solution concepts in sequential
games. They are practically modelled by type structures, which allow the
analyst to represent the players' hierarchies without specifying an infinite
sequence of conditional beliefs. Here, we study type structures that satisfy
a \textquotedblleft richness\textquotedblright \ property, called \textit{%
completeness}. This property is defined on the type structure alone, without
explicit reference to hierarchies of beliefs or other type structures. We
provide sufficient conditions under which a complete type structure
represents all hierarchies of conditional beliefs. In particular, we present
an extension of the main result in Friedenberg (2010) to type structures
with conditional beliefs.

\textbf{Keywords: }Conditional Probability Systems, Hierarchies of Conditional Beliefs, Type Structures, Completeness, Terminality.
\end{abstract}

\section{Introduction}

Hierarchies of conditional beliefs (Battigalli and Siniscalchi 1999) play a
central role for the epistemic analysis of solution concepts in sequential
games. Conditional beliefs generalize ordinary probabilistic beliefs in
that every player is endowed with a collection of conditioning events, and
forms conditional beliefs given each hypothesis in a way that updating is
satisfied whenever possible. Such a collection of measures is called
conditional probability system (CPS, hereafter). A player's first-order
conditional beliefs are described by a CPS over the space of primitive
uncertainty; her second-order conditional beliefs are described by a CPS
over the spaces of primitive uncertainty and of the co-players' first-order
conditional beliefs; and so on.

Battigalli and Siniscalchi (1999) show that hierarchies of CPSs can be
practically represented by \textit{conditional type structures}, i.e.,
compact models which mimic Harsanyi's representation of hierarchies of
probabilistic beliefs. Namely, for each agent there is a set of types. Each
type is associated with a CPS over the set of primitive uncertainty and the
set of the co-players' types. Such structure induces a(n infinite) hierarchy
of CPSs for each type.

Here, we study conditional type structures that satisfy a \textquotedblleft
richness\textquotedblright \ property, called \textit{completeness}. This
property---which plays a crucial role for epistemic foundations\footnote{%
See Dekel and Siniscalchi (2015) for a survey.} of some solution
concepts---is defined on the type structure alone, without explicit
reference to hierarchies of CPSs or other type structures. Loosely speaking,
a type structure is complete if it induces all possible conditional beliefs
about types.

We ask: When does a complete type structure represent all hierarchies of
conditional beliefs? The main result of the paper (Theorem 1) can be briefly
summarized as follows. Suppose that a (conditional) type structure is
complete. Then:

(i) if the structure is Souslin, then it is \textit{finitely terminal},
i.e., it induces all finite order conditional beliefs;

(ii) if the structure is compact and continuous, then it is \textit{terminal}%
, i.e., it induces all hierarchies of conditional beliefs.

Precise definitions are given in the main text. Here we point out that
Theorem 1 is an extension of the main result in Friedenberg (2010) to
conditional type structures. Specifically, Friedenberg studies complete type
structures with beliefs represented by ordinary probabilities; her main
result shows that (i) and (ii) are sufficient conditions for finite
terminality and terminality, respectively. Friedenberg (2010, Section 5)
leaves open the question whether her result still holds when beliefs are
represented by CPSs: our result provides an affirmative answer.

To prove the main result of the paper (Theorem 1), we adopt an approach that
is different from the one in Friedenberg (2010). Specifically, we provide a\
construction---based on the set-up in Heifetz (1993)---of the canonical
space of hierarchies of CPSs which allows us to characterize the notion of
(finite) terminality in a convenient way (Proposition 2). With this, the
crucial step of the proof relies on Lemma 3, an \textquotedblleft
extension\textquotedblright \ result for CPSs whose proof makes use of a
selection argument. The details are spelled out in the paper.\footnote{%
The paper can be found at https://arxiv.org/abs/2305.08940.}

\section{Preliminaries}

A measurable space is a pair $\left( X,\Sigma _{X}\right) $, where $X$ is a
set and $\Sigma _{X}$ is a $\sigma $-algebra, the elements of which are
called \textbf{events}. Throughout this paper, when it is clear from the
context which $\sigma $-algebra on $X$ we are considering, we suppress
reference to $\Sigma _{X}$ and simply write $X$\ to denote a measurable
space. Furthermore, given a function $f:X\rightarrow Y$\ and a family $%
\mathcal{F}_{Y}$\ of subsets of $Y$, we let%
\begin{equation*}
f^{-1}\left( \mathcal{F}_{Y}\right) :=\left \{ E\subseteq X:\exists F\in 
\mathcal{F}_{Y},E=f^{-1}\left( F\right) \right \} \text{.}
\end{equation*}%
So, if $Y$\ is a measurable space, then $f^{-1}\left( \Sigma _{Y}\right) $\
is the $\sigma $-algebra on $X$\ generated by $f$.

We write $\Delta \left( X\right) $\ for the set of probability measures on $%
\Sigma _{X}$. Fix measurable spaces $X$ and $Y$. Given a measurable function 
$f:X\rightarrow Y$, we let $\mathcal{L}_{f}:\Delta \left( X\right)
\rightarrow \Delta \left( Y\right) $ denote the pushforward-measure map
induced by $f$; that is, for each $\mu \in \Delta \left( X\right) $, $%
\mathcal{L}_{f}\left( \mu \right) $\ is the image measure of $\mu $\ under $%
f $, and is defined by $\mathcal{L}_{f}\left( \mu \right) \left( E\right)
:=\mu \left( f^{-1}\left( E\right) \right) $\ for every $E\in \Sigma _{Y}$.

If $X$ is a topological space, we keep using $\Sigma _{X}$\ to denote the
Borel $\sigma $-algebra on $X$. All the topological spaces considered in
this paper are assumed to be metrizable. We consider any product, finite or
countable, of metrizable spaces as a metrizable space with the product
topology. Moreover, we endow each subset of a metrizable space with the
subspace topology. A \textbf{Souslin}\ (resp. \textbf{Lusin}) \textbf{space}
is a topological space that is the image of a complete, separable metric
space under a continuous surjection (resp. bijection). Clearly, a Lusin
space is also Souslin. Examples of Souslin (resp. Lusin) spaces include
analytic (resp. Borel) subsets of a complete separable metric space. In
particular, a Polish space (i.e., a topological space which is homeomorphic
to a complete, separable metric space) is a Lusin space. Furthermore, if $X$
is a Lusin space, then $\left( X,\Sigma _{X}\right) $\ is a \textbf{standard
Borel} space, i.e., there is a Polish space $Y$\ such that $\left( X,\Sigma
_{X}\right) $\ is isomorphic to $\left( Y,\Sigma _{Y}\right) $. If $X$ is a
Souslin space, then $\left( X,\Sigma _{X}\right) $\ is an \textbf{analytic
measurable }space, i.e., there is a Polish space $Y$\ and an analytic subset 
$A\subseteq Y$ such that $\left( X,\Sigma _{X}\right) $\ is isomorphic to $%
\left( A,\Sigma _{A}\right) $; see Cohn (2013, Chapter 8).

For a metrizable space $X$, the set $\Delta \left( X\right) $ of (Borel)
probability measures is endowed with the topology of weak convergence. With
this topology, $\Delta \left( X\right) $\ becomes a metrizable space.

\section{Conditional probability systems}

We represent the players' beliefs as conditional probability systems (cf. R%
\'{e}nyi 1955). Fix a measurable space $\left( X,\Sigma _{X}\right) $. A
family of \textbf{conditioning events} of $X$ is a non-empty family $%
\mathcal{B}\subseteq \Sigma _{X}$ that does not include the empty set. A
possible interpretation is that an individual is uncertain about the
realization of the \textquotedblleft state\textquotedblright \ $x\in X$, and 
$\mathcal{B}$\ represents a family of observable events or \textquotedblleft
relevant hypotheses.\textquotedblright \ If $X$ is a metrizable space, then
each conditioning event $B\in \mathcal{B}$\ is a Borel subset of $X$. In
this case, we say that $\mathcal{B}$ is \textbf{clopen} if each element $%
B\in \mathcal{B}$ is both closed and open. For instance, $\mathcal{B}$ is
clopen if $X$ is a (finite) set endowed with the discrete topology; if $%
\mathcal{B=}\left \{ X\right \} $, then $\mathcal{B}$\ is trivially clopen.

\begin{definition}
\label{Definition CPS}Let $\left( X,\Sigma _{X}\right) $\ be a measurable
space and $\mathcal{B}\subseteq \Sigma _{X}$ be a family\ of conditioning
events. A \textbf{conditional probability system} (\textbf{CPS}) on $\left(
X,\Sigma _{X},\mathcal{B}\right) $\ is an array of probability measures $\mu
:=\left( \mu \left( \cdot |B\right) \right) _{B\in \mathcal{B}}$ such that:%
\newline
\  \ (i) for all $B\in \mathcal{B}$, $\mu \left( B|B\right) =1$;\newline
\  \ (ii) for all $A\in \Sigma _{X}$ and $B,C\in \mathcal{B}$, if $A\subseteq
B\subseteq C$ then $\mu \left( A|B\right) \mu \left( B|C\right) =\mu \left(
A|C\right) $.
\end{definition}

Definition 1 says that a CPS $\mu $ is an element of the set $\Delta \left(
X\right) ^{\mathcal{B}}$, i.e., $\mu $\ is a function from $\mathcal{B}$\ to 
$\Delta \left( X\right) $.\footnote{%
For every pair of sets $X$ and $Y$, we let $Y^{X}$ denote the set of
functions with domain $X\,$\ and codomain $Y$.} We write $\mu \left( \cdot
|B\right) $ to stress the interpretation as a conditional probability given
event $B\in \mathcal{B}$. Condition (ii) is the \textbf{chain rule} of
conditional probabilities and it can be written as follows: if $A\subseteq
B\subseteq C$, then%
\begin{equation*}
\mu \left( B|C\right) >0\Rightarrow \mu \left( A|B\right) =\frac{\mu \left(
A|C\right) }{\mu \left( B|C\right) }\text{.}
\end{equation*}

We let $\Delta ^{\mathcal{B}}\left( X\right) $\ denote the set of CPSs on $%
\left( X,\Sigma _{X},\mathcal{B}\right) $. The following result (whose proof
can be found in Appendix A) records some topological properties of $\Delta ^{%
\mathcal{B}}\left( X\right) $\ when $X$ is a metrizable space and $\mathcal{B%
}$\ is countable.\footnote{%
Lemma 1\ is a generalization of analogous results (for the case when $X$ is
Polish) in Battigalli and Siniscalchi (1999, Lemma 1).}

\begin{lemma}
Fix a metrizable space $X$\ and a countable family $\mathcal{B}\subseteq
\Sigma _{X}$ of conditioning events.\newline
\  \ (i) The space $\Delta ^{\mathcal{B}}\left( X\right) $\ is metrizable.%
\newline
\  \ (ii) If $X$ is Souslin or Lusin, so is $\Delta ^{\mathcal{B}}\left(
X\right) $. \newline
\  \ (iii) Suppose that $\mathcal{B}$\ is clopen. Then $\Delta ^{\mathcal{B}%
}\left( X\right) $ is compact if and only if $X$ is compact.
\end{lemma}

Note that if $X$ is a Polish space, then $\Delta ^{\mathcal{B}}\left(
X\right) $ may fail to be Polish. But, by Lemma 1.(ii), $\Delta ^{\mathcal{B}%
}\left( X\right) $\ is a Lusin space. We can conclude that $\Delta ^{%
\mathcal{B}}\left( X\right) $\ is a Polish space\ provided that $\mathcal{B}$%
\ is clopen (cf. Battigalli and Siniscalchi 1999, Lemma 1).

Fix measurable spaces $\left( X,\Sigma _{X}\right) $\ and $\left( Y,\Sigma
_{Y}\right) $, and families $\mathcal{B}_{X}\subseteq \Sigma _{X}$\ and $%
\mathcal{B}_{Y}\subseteq \Sigma _{Y}$ of conditioning events. Suppose that $%
f:X\rightarrow Y$\ is a measurable function such that%
\begin{equation*}
f^{-1}\left( \mathcal{B}_{Y}\right) =\mathcal{B}_{X}\text{.}
\end{equation*}%
The function $\overline{\mathcal{L}}_{f}:\Delta ^{\mathcal{B}_{X}}\left(
X\right) \rightarrow \Delta ^{\mathcal{B}_{Y}}\left( Y\right) $ defined by%
\begin{equation*}
\overline{\mathcal{L}}_{f}\left( \mu \right) \left( E|B\right) :=\mu \left(
f^{-1}\left( E\right) |f^{-1}\left( B\right) \right) \text{,}
\end{equation*}%
where $E\in \Sigma _{Y}$\ and $B\in \mathcal{B}_{Y}$, is the \textbf{%
pushforward-CPS map} induced by $f$. Note that, for any $\mu \in
\Delta ^{\mathcal{B}_{X}}\left( X\right) $, we can write $\overline{\mathcal{%
L}}_{f}\left( \mu \right) $\ as%
\begin{equation*}
\overline{\mathcal{L}}_{f}\left( \mu \right) =\left( \mathcal{L}_{f}\left(
\mu \left( \cdot |f^{-1}\left( B\right) \right) \right) \right) _{B\in 
\mathcal{B}_{Y}}\text{,}
\end{equation*}%
where $\mathcal{L}_{f}:\Delta \left( X\right) \rightarrow \Delta \left(
Y\right) $\ is the pushforward-measure map induced by $f$.

We record some basic results on the pushforward-CPS map that will be used
extensively throughout the paper. In particular, Lemma 2.(i) ensures that $%
\overline{\mathcal{L}}_{f}$\ is well-defined and justifies the terminology:
if $\mu \in \Delta ^{\mathcal{B}_{X}}\left( X\right) $, then $\overline{%
\mathcal{L}}_{f}\left( \mu \right) $\ is a CPS on $\left( Y,\Sigma _{Y},%
\mathcal{B}_{Y}\right) $.

\begin{lemma}
Fix measurable spaces $\left( X,\Sigma _{X}\right) $\ and $\left( Y,\Sigma
_{Y}\right) $, and families $\mathcal{B}_{X}\subseteq \Sigma _{X}$ and $%
\mathcal{B}_{Y}\subseteq \Sigma _{Y}$\ of conditioning events. Suppose that $%
f:X\rightarrow Y$\ is a measurable function such that $f^{-1}\left( \mathcal{%
B}_{Y}\right) =\mathcal{B}_{X}$. The following statements hold.\newline
\  \ (i) The map $\overline{\mathcal{L}}_{f}:\Delta ^{\mathcal{B}_{X}}\left(
X\right) \rightarrow \Delta ^{\mathcal{B}_{Y}}\left( Y\right) $\ is
well-defined.\newline
\  \ (ii) Suppose that $\mathcal{B}_{X}$\ and $\mathcal{B}_{Y}$\ are
countable, $X$\ is a metrizable space and $Y$ is a Souslin space. If $f$ is
Borel measurable (resp. continuous), then $\overline{\mathcal{L}}_{f}$\ is
Borel measurable (resp. continuous).
\end{lemma}

A special case of image CPS induced by a function is of particular
interest---namely, the marginalization of a CPS on a product space. Consider
measurable spaces $X$ and $Y$, and denote by $\pi _{X}$\ the coordinate
projection from $X\times Y$\ onto $X$. Fix a family $\mathcal{B}\subseteq
\Sigma _{X}$ of conditioning events, and define $\mathcal{B}_{X\times Y}$\ as%
\begin{equation}
\mathcal{B}_{X\times Y}:=\left( \pi _{X}\right) ^{-1}\left( \mathcal{B}%
\right) =\left \{ C\subseteq X\times Y:\exists B\in \mathcal{B},C=B\times
Y\right \} \text{,}
\end{equation}%
that is, $\mathcal{B}_{X\times Y}$\ is the set of all cylinders $B\times Y$\
with $B\in \mathcal{B}$. The function $\overline{\mathcal{L}}_{\pi
_{X}}:\Delta ^{\mathcal{B}_{X\times Y}}\left( X\times Y\right) \rightarrow
\Delta ^{\mathcal{B}}\left( X\right) $ defined by%
\begin{equation*}
\overline{\mathcal{L}}_{\pi _{X}}\left( \mu \right) :=\left( \mathcal{L}%
_{\pi _{X}}\left( \mu \left( \cdot |B\right) \right) \right) _{B\in \mathcal{%
B}}
\end{equation*}%
is called \textbf{marginal-CPS map}, and $\overline{\mathcal{L}}_{\pi
_{X}}\left( \mu \right) $ is called the \textbf{marginal}\ on $X$ of $\mu
\in \Delta ^{\mathcal{B}_{X\times Y}}\left( X\times Y\right) $.

\section{Type structures and hierarchies of conditional beliefs \label%
{Section of type structures and belief hierarchies}}

Throughout, we fix a two-player set $I$;\footnote{%
The assumption of a two-player set is merely for notational convenience. The
analysis can be equivalently carried out with any finite set $I$ with
cardinality greater than two.} given a player $i\in I$, we let $j$ denote
the other player in $I$. We assume that both players share a common
measurable space $\left( S,\Sigma _{S}\right) $, called \textbf{space of
primitive uncertainty}\textit{. }For each $i\in I$, there is a family $%
\mathcal{B}_{i}\subseteq \Sigma _{S}$ of conditioning events. One
interpretation (which is borrowed from Battigalli and De Vito 2021) is the
following: $S$ is a product set, viz. $S:=\times _{i\in I}S_{i}$, and each
element $s:=\left( s_{i}\right) _{i\in I}$\ is an objective description of
players' behavior in a game with complete information and without chance
moves---technically, $\left( s_{i}\right) _{i\in I}$\ is a strategy profile.
Each player is uncertain about the \textquotedblleft true\textquotedblright \
behavior $s\in S$, including his own. If the game has sequential moves, then
each $\mathcal{B}_{i}$\ is a collection of \textit{observable events}; that
is, each $B\in \mathcal{B}_{i}$\ is the set of strategy profiles inducing an
information set of player $i$. Other interpretations of $S$ and $\left( 
\mathcal{B}_{i}\right) _{i\in I}$ are also possible; a more thorough
discussion can be found in Battigalli and Siniscalchi (1999, pp. 191-192).
The results in this paper do not hinge on a specific interpretation.

From now on, we maintain the following technical assumptions on $S$\ and $%
\left( \mathcal{B}_{i}\right) _{i\in I}$:

\begin{itemize}
\item $S$ is a Souslin space, and

\item $\mathcal{B}_{i}\subseteq \Sigma _{S}$\ is countable for every $i\in I$%
.
\end{itemize}

Following Battigalli and Siniscalchi (1999), we adopt the following
notational convention.

\bigskip

\noindent \textbf{Convention 1}. Given a product space $X\times Y$ and a
family $\mathcal{B}\subseteq \Sigma _{X}$ of conditioning events of $X$, the
family of conditioning events of $X\times Y$ is $\mathcal{B}_{X\times Y}$ as
defined in (3.1). Accordingly, we let $\Delta ^{\mathcal{B}}\left( X\times
Y\right) $\ denote the set of CPSs on $\left( X\times Y,\Sigma _{X\times Y},%
\mathcal{B}_{X\times Y}\right) $.

\subsection{Type structures}

We use the framework of type structures (or \textquotedblleft type
spaces\textquotedblright ) to model players' hierarchies of conditional
beliefs. We adopt the following definition of type structure (cf. Battigalli
and Siniscalchi 1999).

\begin{definition}
An $\left( S,\left( \mathcal{B}_{i}\right) _{i\in I}\right) $-\textbf{based
type structure} is a tuple 
\begin{equation*}
\mathcal{T}:=\left( S,\left( \mathcal{B}_{i},T_{i},\beta _{i}\right) _{i\in
I}\right)
\end{equation*}%
such that, for every $i\in I$,\newline
\  \ (i) the \textbf{type set} $T_{i}$\ is a metrizable space;\newline
\  \ (ii) the \textbf{belief map} $\beta _{i}:T_{i}\rightarrow \Delta ^{%
\mathcal{B}_{i}}\left( S\times T_{j}\right) $ is Borel measurable.

Each element of $T_{i}$, viz. $t_{i}$, is called (player $i$'s) \textbf{type}%
.
\end{definition}

Definition 2\ says that, for any $i\in I$, $T_{i}$ represents the set of
player $i$'s possible \textquotedblleft ways to think.\textquotedblright \
Each type $t_{i}\in T_{i}$ is associated with a CPS on the set of primitive
uncertainty as well as on the possible \textquotedblleft ways to
think\textquotedblright \ (types)\ of player $j$. Each conditioning event
for $\beta _{i}\left( t_{i}\right) $ has the form $B\times T_{j}$ with $B\in 
\mathcal{B}_{i}$.

If $\mathcal{B}_{i}\mathcal{=}\left \{ S\right \} $ for every player $i\in I$%
, then each set $\Delta ^{\mathcal{B}_{i}}\left( S\times T_{j}\right) $\ can
be naturally identified with $\Delta \left( S\times T_{j}\right) $. In this
case, Definition 2\ coincides essentially with the definition in Friedenberg
(2010),\footnote{%
The only difference is that $S$ is assumed to be a Polish space in
Friedenberg (2010). Such difference is immaterial for the remainder of the
analysis.} and we say that $\mathcal{T}$\ is an \textbf{ordinary type
structure}. Moreover, we will sometimes refer to type structures via
Definition 2\ as \textbf{conditional type structures}.

\begin{definition}
An $\left( S,\left( \mathcal{B}_{i}\right) _{i\in I}\right) $-based type
structure $\mathcal{T}:=\left( S,\left( \mathcal{B}_{i},T_{i},\beta
_{i}\right) _{i\in I}\right) $ is\newline
\  \ (i) \textbf{Souslin} (resp. \textbf{Lusin}, \textbf{compact}) if, for
every $i\in I$, the type set $T_{i}$ is a Souslin (resp. Lusin, compact)
space;\footnote{%
In Friedenberg (2010), a(n ordinary) type structure is called analytic if
each type set is an analytic subset of a Polish space---hence, a metrizable
Souslin (sub)space. We adopt the definition of Souslin type structure
because we can extend our analysis (as we do in the Supplementary Appendix
of the paper) without assuming metrizability of the topological spaces.}%
\newline
\  \ (ii) \textbf{continuous} if, for every $i\in I$, the belief map $\beta
_{i}$\ is continuous.
\end{definition}

Next, we introduce the notion of completeness for a type structure.

\begin{definition}
An $\left( S,\left( \mathcal{B}_{i}\right) _{i\in I}\right) $-based type
structure $\mathcal{T}:=\left( S,\left( \mathcal{B}_{i},T_{i},\beta
_{i}\right) _{i\in I}\right) $ is \textbf{complete} if, for every $i\in I$,
the belief map $\beta _{i}$ is surjective.
\end{definition}

In words, completeness says that, for each player $i$, and for each
conditional belief $\mu \in \Delta ^{\mathcal{B}_{i}}\left( S\times
T_{j}\right) $\ that player $i$ can hold, there is a type of player $i$\
which induces that belief. Thus, it is a \textquotedblleft
richness\textquotedblright \ requirement which may not be satisfied by some
type structures. For instance, suppose that $S$\ is not a singleton.\ Then a
type structure\ where the type set of some player\ has finite cardinality is
not complete.

A type structure provides an implicit representation of the hierarchies of
beliefs. To address the question whether a complete type structure
represents all hierarchies of beliefs, we need to formally clarify \textit{%
how} type structures generate a collection of hierarchies of beliefs for
each player. This is illustrated in the following section.

\subsection{The canonical space of hierarchies}

In this section we first offer a construction of the set of all hierarchies
of conditional beliefs satisfying a \textit{coherence} condition. Loosely
speaking, coherence means that lower-order beliefs are the marginals of
higher-order beliefs. The construction---which is based on the set-up in
Heifetz (1993)---shows that this set of hierarchies identifies in a natural
way a type structure, which we call it \textquotedblleft
canonical.\textquotedblright \ Next, we show how each profile of types in a
type structure can be associated with an element of the constructed set of
hierarchies. This part is standard (cf. Heifetz and Samet 1998).

\subsubsection{From hierarchies to types}

To construct the set of hierarchies of conditional beliefs, we define
recursively, for each player, two sequences of sets as well as a sequence of
conditioning events. The first sequence, $\left( \Theta _{i}^{n}\right)
_{n\geq 0}$, represents player $i$'s $\left( n+1\right) $-order domain of
uncertainty, for each $n\geq 0$. The second sequence, $\left(
H_{i}^{n}\right) _{n\geq 1}$, consists of player $i$'s $n$-tuples of \textit{%
coherent} conditional beliefs over these space. The notion of coherence,
formally defined below, says that, conditional on any relevant hypothesis,
beliefs at different order do not contradict one another.

Formally, for each player $i\in I$,\ let%
\begin{eqnarray*}
\Theta _{i}^{0} &:&=S\text{,} \\
\mathcal{B}_{i}^{0} &:&=\mathcal{B}_{i}\text{,} \\
H_{i}^{1} &:&=\Delta ^{\mathcal{B}_{i}^{0}}\left( \Theta _{i}^{0}\right) 
\text{.}
\end{eqnarray*}%
The set $\Theta _{i}^{0}$ is player $i$'s $1$-order (primitive) domain of
uncertainty, and a first-order belief, viz. $\mu _{i}^{1}$, is an element of
the set $H_{i}^{1}$.

For $n\geq 1$, assume that $\left( \Theta _{i}^{m}\right) _{m=0,...,n-1}$, $%
\left( \mathcal{B}_{i}^{m}\right) _{m=0,...,n-1}$ and $\left(
H_{i}^{m}\right) _{m=1,...,n}$\ have been defined for each player $i\in I$.
Then, for each $i\in I$,\ let%
\begin{equation*}
\Theta _{i}^{n}:=\Theta _{i}^{0}\times H_{j}^{n}\text{.}
\end{equation*}%
That is, $\Theta _{i}^{n}$\ is player $i$'s $\left( n+1\right) $-order
domain of uncertainty: it consists of the space of primitive uncertainty and
what player $j\neq i$ believes about the space of primitive uncertainty,
what player $j$ believes about what player $i$ believes about the space of
primitive uncertainty,..., and so on, up to level $n$. For each $i\in I$ and 
$n\geq 1$, let $\pi _{i}^{n,n+1}:H_{i}^{n+1}\rightarrow H_{i}^{n}$\ and $%
\rho _{i}^{n-1,n}:\Theta _{i}^{n}\rightarrow \Theta _{i}^{n-1}$ denote the
coordinate projections. By construction, these maps satisfy the following
property:%
\begin{equation*}
\forall i\in I,\forall n\geq 2,\rho _{i}^{n-1,n}=\left( \mathrm{Id}_{\Theta
_{i}^{0}},\pi _{j}^{n-1,n}\right) \text{,}
\end{equation*}%
where $\mathrm{Id}_{\Theta _{i}^{0}}$\ is the identity on $\Theta _{i}^{0}$.

To define players' conditional beliefs on the $\left( n+1\right) $-th order
domain of uncertainty, for each player $i\in I$, let%
\begin{eqnarray*}
\mathcal{B}_{i}^{n} &:&=\left( \rho _{i}^{n-1,n}\right) ^{-1}\left( \mathcal{%
B}_{i}^{n-1}\right) =\left \{ C\subseteq \Theta _{i}^{n}:\exists B\in 
\mathcal{B}_{i}^{n-1},C=\left( \rho _{i}^{n-1,n}\right) ^{-1}\left( B\right)
\right \} \text{,} \\
H_{i}^{n+1} &:&=\left \{ \left( \left( \mu _{i}^{1},...,\mu _{i}^{n}\right)
,\mu _{i}^{n+1}\right) \in H_{i}^{n}\times \Delta ^{\mathcal{B}%
_{i}^{n}}\left( \Theta _{i}^{n}\right) :\overline{\mathcal{L}}_{\rho
_{i}^{n-1,n}}\left( \mu _{i}^{n+1}\right) =\mu _{i}^{n}\right \} \text{.}
\end{eqnarray*}%
Specifically, $\mathcal{B}_{i}^{n}$\ represents the set of relevant
hypotheses upon which player $i$'s $\left( n+1\right) $-th order conditional
beliefs are defined. That is, $\mu _{i}^{n+1}\in \Delta ^{\mathcal{B}%
_{i}^{n}}\left( \Theta _{i}^{n}\right) $ is player $i$'s $\left( n+1\right) $%
-th order CPS with $\mu _{i}^{n+1}\left( \cdot |B\right) \in \Delta \left(
\Theta _{i}^{n}\right) $, $B\in \mathcal{B}_{i}^{n}$. Recursively, it can be
checked that, for all $n\geq 1$,%
\begin{equation*}
\mathcal{B}_{i}^{n}=\left \{ C\subseteq \Theta _{i}^{n}:\exists B\in 
\mathcal{B}_{i},C=B\times H_{j}^{n}\right \} \text{,}
\end{equation*}%
i.e.,\ $\mathcal{B}_{i}^{n}$\ is a set of cylinders in $\Theta _{i}^{n}$\
generated by $\mathcal{B}_{i}$. If $\mathcal{B}_{i}$\ is clopen, then every $%
B\in \mathcal{B}_{i}^{n}$\ is clopen in $\Theta _{i}^{n}$, since each
coordinate projection $\rho _{i}^{n-1,n}$\ is a continuous function. By
definition of each $\Theta _{i}^{n}$, we write, according to Convention 1,%
\begin{equation*}
\Delta ^{\mathcal{B}_{i}^{n}}\left( \Theta _{i}^{n}\right) =\Delta ^{%
\mathcal{B}_{i}}\left( \Theta _{i}^{n}\right) \text{.}
\end{equation*}

The set $H_{i}^{n+1}$\ is the set of player $i$'s $\left( n+1\right) $%
-tuples of CPSs on $\Theta _{i}^{0}$, $\Theta _{i}^{1}$,..., $\Theta
_{i}^{n} $. The condition on $\mu _{i}^{n+1}$\ in the definition of $%
H_{i}^{n+1}$\ is the \textbf{coherence} condition mentioned above. Given the
recursive construction of the sets, CPSs $\mu _{i}^{n+1}$\ and $\mu _{i}^{n}$%
\ both specify a (countable) array of conditional beliefs on the domain of
uncertainty $\Theta _{i}^{n-1}$, and those beliefs cannot be contradictory.
Formally, for all $B\in \mathcal{B}_{i}^{n-1}$ and for every event $%
E\subseteq \Theta _{i}^{n-1}$,%
\begin{equation*}
\mu _{i}^{n+1}\left( \left( \rho _{i}^{n-1,n}\right) ^{-1}\left( E\right)
\left \vert \left( \rho _{i}^{n-1,n}\right) ^{-1}\left( B\right) \right.
\right) =\mu _{i}^{n}\left( E|B\right) \text{.}
\end{equation*}%
That is, the conditional belief $\mu _{i}^{n+1}(\cdot |(\rho
_{i}^{n-1,n})^{-1}\left( B\right) )$\ must assign to event $\left( \rho
_{i}^{n-1,n}\right) ^{-1}\left( E\right) $\ the same number as $\mu
_{i}^{n}\left( \cdot |B\right) $\ assigns to event $E$.

\begin{remark}
For each $i\in I$\ and $n\geq 1$, the set $H_{i}^{n+1}$\ is a closed subset
of $H_{i}^{n}\times \Delta ^{\mathcal{B}_{i}}\left( \Theta _{i}^{n}\right) $%
. So $H_{i}^{n+1}$ is a Souslin (resp. Lusin) space provided $S$ is a
Souslin (resp. Lusin) space. If $\mathcal{B}_{i}$ is clopen for each $i\in I$%
, then $H_{i}^{n+1}$ is compact if and only if $S$ is compact.
\end{remark}

In the limit, for each $i\in I$, let%
\begin{align*}
H_{i}& :=\left \{ \left( \mu _{i}^{1},\mu _{i}^{2},...\right) \in \times
_{n=0}^{\infty }\Delta ^{\mathcal{B}_{i}}\left( \Theta _{i}^{n}\right)
:\forall n\geq 1,\left( \mu _{i}^{1},...,\mu _{i}^{n}\right) \in
H_{i}^{n}\right \} \text{,} \\
\Theta _{i}& :=S\times H_{j}\text{.}
\end{align*}

\begin{remark}
The set $H_{i}$\ is a closed subset of $\times _{n=0}^{\infty }\Delta ^{%
\mathcal{B}_{i}}\left( \Theta _{i}^{n}\right) $. So $H_{i}$ is a Souslin
(resp. Lusin) space provided $S$ is a Souslin (resp. Lusin) space. If $%
\mathcal{B}_{i}$ is clopen for each $i\in I$, then $H_{i}$ is compact if and
only if $S$ is compact.
\end{remark}

The following result corresponds to Proposition 2 in Battigalli and
Siniscalchi (1999).

\begin{proposition}
For each $i\in I$, the spaces $H_{i}$\ and $\Delta ^{\mathcal{B}_{i}}\left(
S\times H_{j}\right) $\ are homeomorphic.
\end{proposition}

The set $H:=\times _{i\in I}H_{i}$ is the set of all pairs of \textit{%
collectively coherent} hierarchies of conditional beliefs; that is, $H$\ is
the set of pairs of coherent hierarchies satisfying common full belief of
coherence.\footnote{%
An event $E$ is fully believed under a CPS $\left( \mu \left( \cdot
|B\right) \right) _{B\in \mathcal{B}}$ \ if $\mu \left( E|B\right) =1$ for
every $B\in \mathcal{B}$. The notion of \textquotedblleft common full belief
of coherence\textquotedblright \ is made explicit in the alternative\
construction of the canonical space \textit{\`{a} la} Battigalli and
Siniscalchi (1999). A note on terminology: in Battigalli and Siniscalchi
(1999) the expression \textquotedblleft certainty\textquotedblright \ is
used in place of \textquotedblleft full belief.\textquotedblright}

The homeomorphisms in Proposition 1 are \textquotedblleft
canonical\textquotedblright \ in the following sense: every coherent
hierarchy $\left( \mu _{i}^{1},\mu _{i}^{2},...\right) $\ of player $i$ is
associated with a unique CPS $\mu _{i}$\ on the space of primitive
uncertainty and the coherent hierarchies of the co-player, i.e., $S\times
H_{j}$. Then, for all $n\geq 0$, the marginal of $\mu _{i}$\ on player $i$'s 
$\left( n+1\right) $-order domain of uncertainty, viz. $\Theta _{i}^{n}$, is
precisely what it should be, namely $\mu _{i}^{n+1}$. A formal definition of
such homeomorphisms is not needed for the statements and proofs of the
results in this paper. Instead, we will make use of the following
implication of Proposition 1: we can define an $\left( S,\left( \mathcal{B}%
_{i}\right) _{i\in I}\right) $-based type structure $\mathcal{T}^{\mathrm{c}%
}:=\left( S,\left( \mathcal{B}_{i},T_{i}^{\mathrm{c}},\beta _{i}^{\mathrm{c}%
}\right) _{i\in I}\right) $\ by letting, for each $i\in I$,%
\begin{equation*}
T_{i}^{\mathrm{c}}:=H_{i}\text{,}
\end{equation*}%
and $\beta _{i}^{\mathrm{c}}:T_{i}^{\mathrm{c}}\rightarrow \Delta ^{\mathcal{%
B}_{i}}\left( S\times T_{j}^{\mathrm{c}}\right) $\ is the \textquotedblleft
canonical\textquotedblright \ homeomorphism. Following the terminology in
the literature, we call $\mathcal{T}^{\mathrm{c}}$\ the \textbf{canonical} 
\textbf{type structure}.\footnote{%
There are differences, in terms of technical assumptions, between our
construction of $\mathcal{T}^{\mathrm{c}}$\ and the one in Battigalli and
Siniscalchi (1999). Actually, our construction is a generalization of theirs
since we allow for weaker conditions on (\textit{i}) the family of
conditioning events, and (\textit{ii}) the topological property of the
primitive uncertainty space $S$.}

\begin{remark}
Structure $\mathcal{T}^{\mathrm{c}}$\ is Souslin, continuous and complete.
If $S$\ is Lusin, then $\mathcal{T}^{\mathrm{c}}$\ is Lusin. If $\mathcal{B}%
_{i}$\ is clopen for each $i\in I$, then $\mathcal{T}^{\mathrm{c}}$\ is a
compact type structure if and only if $S$\ is a compact space.
\end{remark}

\subsubsection{From types to hierarchies}

The next step is to consider the relationship between the set of hierarchies
constructed in the previous section and any other type structure. In so
doing, we specify how types generate (collectively) coherent hierarchies of
conditional beliefs. As we did in the previous section, given a set $X$, we
let $\mathrm{Id}_{X}$\ denote\ the identity map.

Fix an $\left( S,\left( \mathcal{B}_{i}\right) _{i\in I}\right) $-based type
structure $\mathcal{T}:=\left( S,\left( \mathcal{B}_{i},T_{i},\beta
_{i}\right) _{i\in I}\right) $.\ We construct a natural (Borel) measurable
map, called \textbf{hierarchy map}, which unfolds the higher-order beliefs
of each player $i\in I$. This map assigns to each $t_{i}\in T_{i}$ a
hierarchy of beliefs in $H_{i}$.

For each $i\in I$, let $h_{-i}^{0}:\Theta _{i}^{0}\times T_{j}\rightarrow
\Theta _{i}^{0}$ be the projection map (recall that $\Theta _{i}^{0}:=S$\
for each $i\in I$). The \textquotedblleft first-order map\textquotedblright
\ for each player $i$, viz. $h_{i}^{1}:T_{i}\rightarrow H_{i}^{1}$, is
defined by%
\begin{equation*}
h_{i}^{1}\left( t_{i}\right) :=\overline{\mathcal{L}}_{h_{-i}^{0}}\left(
\beta _{i}\left( t_{i}\right) \right) \text{.}
\end{equation*}%
In words, $h_{i}^{1}\left( t_{i}\right) $\ is the marginal on $S$\ of CPS $%
\beta _{i}\left( t_{i}\right) $. Measurability of each map $h_{i}^{1}$ holds
by Lemma 2 and measurability of belief maps.

With this, for each $i\in I$, let $h_{-i}^{1}:\Theta _{i}^{0}\times
T_{j}\rightarrow \Theta _{i}^{0}\times H_{j}^{1}=\Theta _{i}^{1}$ be the map
defined as $h_{-i}^{1}:=\left( \mathrm{Id}_{S},h_{j}^{1}\right) $; i.e., for
each pair $\left( s,t_{j}\right) \in \Theta _{i}^{0}\times T_{j}$, the
expression $h_{-i}^{1}\left( s,t_{j}\right) =\left( s,h_{j}^{1}\left(
t_{j}\right) \right) \in \Theta _{i}^{1}$\ describes the profile\ $s$\ and
the first-order beliefs for type $t_{j}\in T_{j}$. Standard arguments show
that, for each $i\in I$, the map $h_{-i}^{1}$\ is measurable; furthermore,
it can be checked that $h_{-i}^{1}$ satisfies 
\begin{equation*}
h_{-i}^{0}=\rho _{i}^{0,1}\circ h_{-i}^{1}\text{ \ and \ }\mathcal{B}%
_{\Theta _{i}^{0}\times T_{j}}=\left( h_{-i}^{1}\right) ^{-1}\left( \mathcal{%
B}_{i}^{1}\right) \text{.}
\end{equation*}

Recursively, we define the \textquotedblleft $\left( n+1\right) $%
th-orders\textquotedblright \ maps. For $n\geq 1$, assume that measurable
maps $h_{i}^{n}:T_{i}\rightarrow H_{i}^{n}$ have been defined for each
player $i\in I$. Moreover, for each $i\in I$, assume that $h_{-i}^{n}:\Theta
_{i}^{0}\times T_{j}\rightarrow \Theta _{i}^{0}\times H_{j}^{n}=\Theta
_{i}^{n}$\ is the unique measurable function, defined as $h_{-i}^{n}:=\left( 
\mathrm{Id}_{S},h_{j}^{n}\right) $, which satisfies%
\begin{equation}
\mathcal{B}_{\Theta _{i}^{0}\times T_{j}}=\left( h_{-i}^{n}\right)
^{-1}\left( \mathcal{B}_{i}^{n}\right)
\label{hierarchy description conditioning events}
\end{equation}%
and%
\begin{equation}
h_{-i}^{n-1}=\rho _{i}^{n-1,n}\circ h_{-i}^{n}\text{.}
\label{Hierarchy description condition}
\end{equation}%
Fix a player $i\in I$. Note that, since (\ref{hierarchy description
conditioning events})\ holds, $\overline{\mathcal{L}}_{h_{-i}^{n}}:\Delta ^{%
\mathcal{B}_{i}}\left( S\times T_{j}\right) \rightarrow \Delta ^{\mathcal{B}%
_{i}^{n}}\left( \Theta _{i}^{n}\right) $ is a well-defined measurable map by
Lemma 2. With this, define $h_{i}^{n+1}:T_{i}\rightarrow H_{i}^{n}\times
\Delta ^{\mathcal{B}_{i}^{n}}\left( \Theta _{i}^{n}\right) $ by%
\begin{equation*}
h_{i}^{n+1}\left( t_{i}\right) :=\left( h_{i}^{n}\left( t_{i}\right) ,%
\overline{\mathcal{L}}_{h_{-i}^{n}}\left( \beta _{i}\left( t_{i}\right)
\right) \right) \text{.}
\end{equation*}%
Using the same arguments as above, it is easily verified that the map $%
h_{i}^{n+1}$ is measurable. By (\ref{Hierarchy description condition}), it
follows that:

\begin{remark}
$h_{i}^{n+1}\left( T_{i}\right) \subseteq H_{i}^{n+1}$\ for each $i\in I$.
\end{remark}

It is easily seen that, for each $t_{i}\in T_{i}$,%
\begin{equation*}
h_{i}^{n+1}\left( t_{i}\right) =\left( \overline{\mathcal{L}}%
_{h_{-i}^{0}}\left( \beta _{i}\left( t_{i}\right) \right) ,...,\overline{%
\mathcal{L}}_{h_{-i}^{n-1}}\left( \beta _{i}\left( t_{i}\right) \right) ,%
\overline{\mathcal{L}}_{h_{-i}^{n}}\left( \beta _{i}\left( t_{i}\right)
\right) \right) \text{.}
\end{equation*}%
Finally, for each $i\in I$, the map $h_{i}:T_{i}\rightarrow \times
_{n=0}^{\infty }\Delta ^{\mathcal{B}_{i}}\left( S\times H_{j}^{n}\right) $\
is defined by%
\begin{equation*}
h_{i}\left( t_{i}\right) :=\left( \overline{\mathcal{L}}_{h_{-i}^{n}}\left(
\beta _{i}\left( t_{i}\right) \right) \right) _{n\geq 0}\text{.}
\end{equation*}%
Thus, $h_{i}\left( t_{i}\right) $\ is the hierarchy generated by type $%
t_{i}\in T_{i}$. Each type generates a (collectively) coherent hierarchy of
beliefs, i.e., $h_{i}\left( T_{i}\right) \subseteq H_{i}$.

\begin{remark}
For each $i\in I$, the map $h_{i}:T_{i}\rightarrow H_{i}$\ is well-defined
and Borel measurable. Furthermore, if $\mathcal{T}$\ is continuous, then,
for each $i\in I$, the map $h_{i}$\ is continuous.
\end{remark}

\section{Terminal type structures}

The following definitions are extensions to conditional type structures of
the definitions put forward by Friedenberg (2010, Section 2) for ordinary
type structures.

\begin{definition}
An $\left( S,\left( \mathcal{B}_{i}\right) _{i\in I}\right) $-based type
structure $\mathcal{T}:=\left( S,\left( \mathcal{B}_{i},T_{i},\beta
_{i}\right) _{i\in I}\right) $\ is \textbf{finitely} \textbf{terminal} if,
for each type structure $\mathcal{T}^{\ast }:=\left( S,\left( \mathcal{B}%
_{i},T_{i}^{\ast },\beta _{i}^{\ast }\right) _{i\in I}\right) $, each type $%
t_{i}^{\ast }\in T_{i}^{\ast }$\ and each $n\in \mathbb{N}$, there is a type 
$t_{i}\in T_{i}$\ such that $h_{i}^{\ast ,n}\left( t_{i}^{\ast }\right)
=h_{i}^{n}\left( t_{i}\right) $.
\end{definition}

\begin{definition}
An $\left( S,\left( \mathcal{B}_{i}\right) _{i\in I}\right) $-based type
structure $\mathcal{T}:=\left( S,\left( \mathcal{B}_{i},T_{i},\beta
_{i}\right) _{i\in I}\right) $\ is \textbf{terminal} if, for each type structure $%
\mathcal{T}^{\ast }:=\left( S,\left( \mathcal{B}_{i},T_{i}^{\ast },\beta
_{i}^{\ast }\right) _{i\in I}\right) $\ and each type $t_{i}^{\ast }\in
T_{i}^{\ast }$, there is a type $t_{i}\in T_{i}$\ such that $h_{i}^{\ast
}\left( t_{i}^{\ast }\right) =h_{i}\left( t_{i}\right) $.
\end{definition}

Definition 5 says that $\mathcal{T}$ is finitely terminal if, for every type 
$t_{i}^{\ast }$ that occurs in some structure $\mathcal{T}^{\ast }$ and
every $n\in \mathbb{N}$, there exists a type $t_{i}$ in $\mathcal{T}$ whose
hierarchy agrees with the hierarchy generated by $t_{i}^{\ast }$\ up to
level $n$. Definition 6 says that $\mathcal{T}$ is terminal if, for every
type $t_{i}^{\ast }$ that occurs in some structure $\mathcal{T}^{\ast }$,
there exists a type $t_{i}$ in $\mathcal{T}$ which generates the same
hierarchy as $t_{i}^{\ast }$.

The notion of terminality in Definition 6 can be equivalently expressed as
follows: $\mathcal{T}$\ is terminal if, for every structure $\mathcal{T}%
^{\ast }$, there exists a \textbf{hierarchy morphism} from $\mathcal{T}%
^{\ast }$ to $\mathcal{T}$, i.e., a map that preserves the hierarchies of
beliefs. Here we show that (a) Definition 5 is equivalent to the requirement
that a type structure generates all finite-order beliefs consistent with
coherence and common full belief of coherence; and (b) Definition 6 is
equivalent to the requirement that a type structure generates all
collectively coherent hierarchies of beliefs.

\begin{remark}
An $\left( S,\left( \mathcal{B}_{i}\right) _{i\in I}\right) $-based type
structure $\mathcal{T}$\ is finitely terminal if and only if, for each $%
\left( S,\left( \mathcal{B}_{i}\right) _{i\in I}\right) $-based type
structure $\mathcal{T}^{\ast }$, each player $i\in I$\ and each $n\in 
\mathbb{N}$,%
\begin{equation*}
h_{i}^{\ast ,n}\left( T_{i}^{\ast }\right) \subseteq h_{i}^{n}\left(
T_{i}\right) \text{.}
\end{equation*}%
An $\left( S,\left( \mathcal{B}_{i}\right) _{i\in I}\right) $-based type
structure $\mathcal{T}$\ is terminal if and only if, for each $\left(
S,\left( \mathcal{B}_{i}\right) _{i\in I}\right) $-based type structure $%
\mathcal{T}^{\ast }$, and for each player $i\in I$,%
\begin{equation*}
h_{i}^{\ast }\left( T_{i}^{\ast }\right) \subseteq h_{i}\left( T_{i}\right) 
\text{.}
\end{equation*}
\end{remark}

The following result establishes the relationship between any (finitely)
terminal type structure and the canonical space of hierarchies.

\begin{proposition}
Fix an $\left( S,\left( \mathcal{B}_{i}\right) _{i\in I}\right) $-based type
structure $\mathcal{T}:=\left( S,\left( \mathcal{B}_{i},T_{i},\beta
_{i}\right) _{i\in I}\right) $.\newline
\  \ (i) $\mathcal{T}$ is finitely terminal if and only if $h_{i}^{n}\left(
T_{i}\right) =H_{i}^{n}$ for each $i\in I$\ and each $n\in \mathbb{N}$.%
\newline
\  \ (ii) $\mathcal{T}$\ is terminal if and only if $h_{i}\left( T_{i}\right)
=H_{i}$ for each $i\in I$.
\end{proposition}

Proposition 2\ provides a characterization of (finite) terminality which
turns out to be useful for the proof of the main result. It is basically a
version of Result 2.1 (and Proposition B1.(ii)) in Friedenberg (2010).

\section{Main result}

The main result of this paper is the following theorem.

\begin{theorem}
Fix an $\left( S,\left( \mathcal{B}_{i}\right) _{i\in I}\right) $-based type
structure $\mathcal{T}:=\left( S,\left( \mathcal{B}_{i},T_{i},\beta
_{i}\right) _{i\in I}\right) $.\newline
\  \ (i) If $\mathcal{T}$\ is\ Souslin and complete, then $\mathcal{T}$ is
finitely terminal.\newline
\  \ (ii) If $\mathcal{T}$\ is complete, compact and continuous, then $%
\mathcal{T}$ is terminal.
\end{theorem}

If $\mathcal{B}_{i}\mathcal{=}\left \{ S\right \} $ for every $i\in I$, then
Theorem 1\ corresponds to Theorem 3.1\ in Friedenberg (2010). The proof of
Theorem 1 relies on the following result, whose proof\ makes use of Von
Neumann Selection Theorem.

\begin{lemma}
\textit{Fix Souslin spaces }$X$\textit{, }$Y$\textit{\ and }$Z$\textit{, and
a countable family }$\mathcal{B}\subseteq \Sigma _{X}$\textit{\ of
conditioning events}. \textit{Let }$f_{1}:Y\rightarrow Z$\textit{\ be Borel
measurable, and define }$f_{2}:X\times Y\rightarrow X\times Z$\textit{\ as }$%
f_{2}:=\left( \mathrm{Id}_{X},f_{1}\right) $\textit{. Then:\newline
\  \ (i) }$f_{2}$\textit{\ is Borel measurable, and the map }$\overline{%
\mathcal{L}}_{f_{2}}:\Delta ^{\mathcal{B}}\left( X\times Y\right)
\rightarrow \Delta ^{\mathcal{B}}\left( X\times Z\right) $\textit{\ is
well-defined;\newline
\  \ (ii) if }$f_{1}$\textit{\ is surjective, then }$\overline{\mathcal{L}}%
_{f_{2}}$\textit{\ is surjective.}
\end{lemma}

The proof of part (i) of Theorem 1 is by induction on $n\in \mathbb{N}$. The
proof of the base step does not rely on the hypothesis that $\mathcal{T}$ is
Souslin. Lemma 3\ is used \textit{only} in the inductive (and crucial) step.
The proof of part (ii) of Theorem 1 uses the same arguments as in
Friedenberg (2010).

Some comments on Theorem 1 are in order. First, complete type structures
that are finitely terminal can be easily constructed. A simple
example---which uses the ideas in Brandenburger et al. (2008, proof of
Proposition 7.2)---is the following. For each $i\in I$, let $T_{i}$ be the
Baire space, i.e., the (non-compact) Polish space $\mathbb{N}^{\mathbb{N}}$.
Every Souslin space is the image of $\mathbb{N}^{\mathbb{N}}$\ under a
continuous function.\footnote{%
It is well-known that every non-empty Polish space is the image of $\mathbb{N%
}^{\mathbb{N}}$\ under a continuous function. Using this result, it is easy
to check---by inspection of definitions---that an analogous conclusion also
holds for Souslin spaces (cf. Cohn 2003, Corollary 8.2.8).} Since $\Delta ^{%
\mathcal{B}_{i}}\left( S\times T_{j}\right) $\ is a Souslin space by Lemma
1, there exists a continuous surjection $\beta _{i}:T_{i}\rightarrow \Delta
^{\mathcal{B}_{i}}\left( S\times T_{j}\right) $. These maps give us a
Souslin and complete type structure $\mathcal{T}$ that is finitely terminal,
but not necessarily terminal. A more complex example of a(n ordinary)
complete, finitely terminal type structure which is \textit{not} terminal
can be found in Friedenberg and Keisler (2021, Section 6).

Second, we point out that complete, compact and continuous type structures
may not exist. This is so because the structural hypothesis on the families
of conditioning events $\mathcal{B}_{i}$ ($i\in I$) are quite weak---each
element of $\mathcal{B}_{i}$\ is a Borel subset of $S$. To elaborate,
suppose that $\mathcal{T}$\ is complete and compact. Completeness yields $%
\beta _{i}\left( T_{i}\right) =\Delta ^{\mathcal{B}_{i}}\left( S\times
T_{j}\right) $\ for each player $i\in I$. If $\mathcal{T}$\ were continuous,
then compactness of $T_{i}$\ would imply compactness of $\Delta ^{\mathcal{B}%
_{i}}\left( S\times T_{j}\right) $\ as well, because the continuous image of
a compact set is compact. But, in general, $\Delta ^{\mathcal{B}_{i}}\left(
S\times T_{j}\right) $\ is not a compact space, even if the underlying space 
$S\times T_{j}$\ is compact (cf. Lemma 1).\footnote{%
Let $X$ be a compact space. The set $\Delta ^{\mathcal{B}}\left( X\right) $\
may fail to be a closed subset of $\Delta \left( X\right) ^{\mathcal{B}}$,
which is a compact metrizable space. Yet, by Lemma 1.(ii), $\Delta ^{%
\mathcal{B}}\left( X\right) $\ is a Lusin subspace of $\Delta \left(
X\right) ^{\mathcal{B}}$. In particular, $\Delta ^{\mathcal{B}}\left(
X\right) $ is a Borel subset of $\Delta \left( X\right) ^{\mathcal{B}}$.} In
other words, unless each family $\mathcal{B}_{i}$ ($i\in I$) satisfies some
specific assumptions, there is no guarantee that a complete, compact and
continuous type structure exists---in particular, complete and continuous
structures may fail the compactness requirement.\footnote{%
Analogously, complete and compact type structures may fail the continuity
requirement.} If this is the case, Theorem 1.(ii) still holds, but \textit{%
vacuously} because the antecedent of the conditional is false. An immediate
implication of this fact is that the completeness test---as formalized by
completeness, compactness and continuity---cannot be applied.

With this in mind, suppose now that each family $\mathcal{B}_{i}$ ($i\in I$%
)\ is clopen. If $S$ is a compact space, then complete, compact and
continuous type structures do exist. The canonical structure is a prominent
example, but there are also complete, compact and continuous structures
which can be distinct from the canonical one. For instance, one can take
each $T_{i}$\ to be the Cantor space $\left \{ 0,1\right \} ^{\mathbb{N}}$,
a compact metrizable space. Lemma 1.(iii) yields that each $\Delta ^{%
\mathcal{B}_{i}}\left( S\times T_{j}\right) $\ is compact metrizable, so
there exists a continuous surjection $\beta _{i}:T_{i}\rightarrow \Delta ^{%
\mathcal{B}_{i}}\left( S\times T_{j}\right) $ (Aliprantis and Border 2006,
Theorem 3.60). The resulting structure $\mathcal{T}$\ is complete, compact
and continuous, hence terminal.

Finally, note that if each $\mathcal{B}_{i}$\ is clopen, then compactness of 
$S$ is a necessary condition for the existence of a complete, compact and
continuous structure $\mathcal{T}$. Indeed, continuity and surjectivity of
the belief maps entail that each set $\Delta ^{\mathcal{B}_{i}}\left(
S\times T_{j}\right) $\ is compact; by Lemma 1.(iii)\ and Tychonoff's
theorem, $S$\ is a compact space.

\bigskip 

\noindent \textbf{Acknowledgements}. I thank three anonymous reviewers from
TARK 2023, Pierpaolo Battigalli and Emiliano Catonini for helpful feedback.
Financial support from the Italian Ministry of Education, PRIN 2017, Grant
Number 2017K8ANN4,\ is gratefully acknowledged.

\nocite{*}
\bibliographystyle{eptcs}
\bibliography{mainpaper}
\end{document}